\documentclass[structabstract]{aa}  
\usepackage{graphicx}
\usepackage{txfonts}
\usepackage{natbib}
\usepackage{color}
\begin{document}
\title{Galactic Bar/Spiral Arm Interactions in NGC\,3627\thanks{Based
    on observations carried out with the IRAM PdBI and 30\,m
    telescope. The data are available in electronic form at the CDS
    via anonymous ftp to cdsarc.u-strasbg.fr (130.79.128.5) or via
    http://cdsweb.u-strasbg.fr/cgi-bin/qcat?J/A+A/}.}


   \author{H.~Beuther
          \inst{1}
          \and
          S.~Meidt
          \inst{1}
          \and
          E.~Schinnerer
          \inst{1}
          \and
          R.~Paladino
           \inst{2}
          \and
          A.~Leroy
          \inst{3}
            }
   \institute{$^1$ Max-Planck-Institute for Astronomy, K\"onigstuhl 17,
              69117 Heidelberg, Germany, \email{name@mpia.de}\\
              $^2$ Dipartimento di Fisica e Astronomia, University of Bologna, Viale Berti Pichat 6/2, 40127, Bologna, Italy; INAF - Istituto di Radioastronomia \& Italian ALMA Regional Centre, via P. Gobetti 101, 40129, Bologna, Italy\\
              $^3$ Department of Astronomy, The Ohio State University, 140 West 18th Avenue, Columbus, OH 43210, USA; National Radio Astronomy Observatory, Charlottesville, VA 22903, USA}

   \date{Version of \today}

\abstract
{}
{To gain insight into the expected gas dynamics at the interface of
  the Galactic bar and spiral arms in our own Milky Way galaxy, we
  examine as an extragalactic counterpart the evidence for
  multiple distinct velocity components in the cold, dense molecular
  gas populating a comparable region at the end of the bar in the
  nearby galaxy NGC 3627. }
{We assemble a high resolution view of molecular gas kinematics traced
  by CO(2-1) emission and extract line-of-sight velocity profiles from
  regions of high and low gas velocity dispersion.}
{The high velocity dispersions arise with often double-peaked or
  multiple line-profiles.  We compare the centroids of the different
  velocity components to expectations based on orbital dynamics in the
  presence of bar and spiral potential perturbations.  A model of the
  region as the interface of two gas-populated orbits families
  supporting the bar and the independently rotating spiral arms
  provides an overall good match to the data. An extent of the bar to
  the corotation radius of the galaxy is favored.}
{Using NGC3627 as an extragalactic example, we expect situations like
  this to favor strong star formation events such as observed in our
  own Milky Way since gas can pile up at the crossings between the
  orbit families.  The relative motions of the material following
  these orbits is likely even more important for the build up of high
  density in the region. The surface densities in NGC3627 are also so
  high that shear at the bar end is unlikely to significantly weaken
  the star formation activity. We speculate that scenarios in which
  the bar and spiral rotate at two different pattern speeds may be the
  most favorable for intense star formation at such interfaces.}


\titlerunning{Bar-spiral interaction in NGC3627}

\maketitle

\section{Introduction}
\label{intro}

The interfaces of galactic bars and outer spiral arms represent some
of most active star-forming environments in the local universe.  In
our own Milky Way, the overlap between the end of the Galactic bar and
the inner Scutum-Centaurus spiral arm at longitudes of $\sim$30
degrees hosts the W43 mini-starburst with an approximate luminosity of
$L\sim 3\times 10^6$\,L$_{\odot}$ (e.g.,
\citealt{blum1999,motte2003,bally2010,nguyen2011}). Questions about
the nature of star formation in such bar-spiral interface regions
remain puzzling, e.g., is it possible that clouds at different
relative velocities physically interact and even may induce the star
formation process this way. \citet{nguyen2011} estimate a star
formation rate (SFR) between 0.01 and 0.1\,M$_{\odot}yr^{-1} \times
\left(\frac{d}{6kpc}\right)^2$ (with d the distance) over the size of
the W43 cloud complex of approximately 15\, (kpc)$^2$. One of the
observational difficulties lies in the fact that gas kinematics are
hard to constrain at these locations, both in our own Milky Way and in
external galaxies.

In the case of W43, several lines of evidence reveal the dynamics of
the bar-spiral interaction and its effect on gas motions as the source
of burst of star formation, which is thought to be the site of
converging flows (e.g.,
\citealt{benjamin2005,lopez2007,rodriguez2008,nguyen2011,carlhoff2013,motte2014}).
This region contains various evolutionary stages, from young infrared
dark clouds to active star-forming cloud and a Wolf-Rayet cluster
\citep{blum1999,beuther2012a}.

A peculiar aspect of the W43 region is the existence of two prominent
gas velocity components along the line of sight, the principle
component at $\sim$100\,km\,s$^{-1}$, and a secondary at
$\sim$50\,km\,s$^{-1}$. \citet{motte2014} discuss several converging
gas flows in that region associated with kinematic CO and HI gas
components between approximately 60 and 120\,km\,s$^{-1}$. In their
picture, the 50\,km\,s$^{-1}$ component is not considered further
  because they do not identify associated converging velocity
  structures towards W43. Similarly, \citet{nguyen2011} and
\citet{carlhoff2013} argue that the more prominent
$\sim$100\,km\,s$^{-1}$ peak is the dominant one for the on-going star
formation in the region, corresponding to the W43 complex, whereas the
$\sim$50\,km\,s$^{-1}$ component may only be a chance alignment that
could be attributed to the Perseus spiral arm. In contrast to
this, \citet{beuther2012a} find both components in the $^{13}$CO(2--1)
emission not just along the same sidelines but the two components
appear in projection spatially connected. Furthermore, both components
are also detected in dense gas tracers like N$_2$H$^+$, and again,
both velocity components appear in mapping studies as connected gas
structures \citep{beuther2012a}. While this may still be a chance
alignment, it is nevertheless suggesting that both components may
emerge from the same region. This part of our Galaxy is also almost
the only region in the Milky Way that exhibits multiple velocity
components in the ionized gas tracer through radio recombination lines
\citep{anderson2011}. This can be interpreted as additional evidence
for potentially interacting gas clouds.

Due to our position in the Galaxy, however, it is currently not
possible to accurately determine the distance of the two components,
and thus to unambiguously determine whether the two gas components in
W43 are either in fact interacting or just chance alignments in
different parts of the Galaxy.  Unambiguous proof will likely depend
on future exact distance measurements via maser parallaxes of the
Bessel project of both velocity components
\citep{brunthaler2011}. While the 100\,km\,s$^{-1}$ component has
recently been determined to be at a distance of $\sim 5.5$\,kpc via
maser parallax measurements \citep{zhang2014}, the comparable
measurement for the 50\,km\,s$^{-1}$ component is still missing.

\begin{figure*}[ht]
\includegraphics[width=0.9\textwidth]{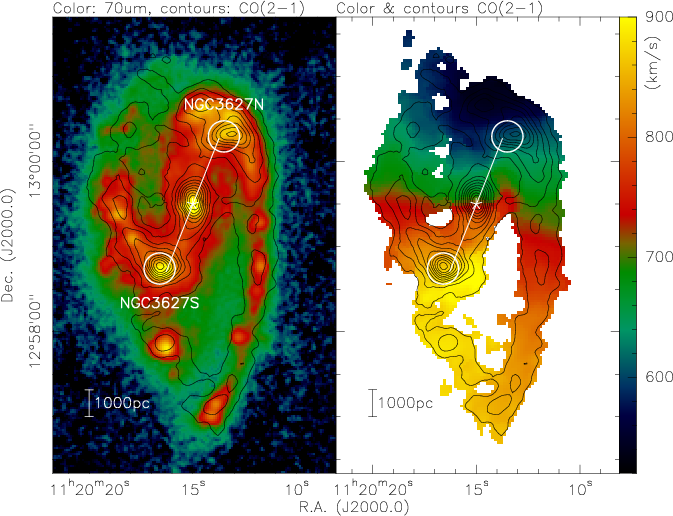}
\caption{Herschel 70\,$\mu$m \citep{kennicutt2011} and HERACLES
  CO(2--1) data of NGC3627 \citep{leroy2009}. The left panel presents
  in color the 70\,$\mu$m emission (logarithmic stretch from 0.0001 to
  0.1\,Jy\,pixel$^{-1}$) while the contours shows the integrated
  CO(2--1) emission from 5 to 55\,K\,km\,s$^{-1}$ in
  5\,K\,km\,s$^{-1}$ steps. In the right panel, the color-scale
  presents the 1st moment map (intensity-weighted velocities), the
  contours again show the integrated CO(2--1) emission.  Circles mark
  the locations and areas of the PdBI(2-1) primary beams for NGC3627N
  and NGC3627S. The central star and white line indicate the bar
  location following \citet{casasola2011}. A linear scale-bar assuming
  11.1\,Mpc distance is shown in the bottom-left corner.}
\label{heracles}
\end{figure*}

Hence, while proofing the spatial (dis)association of velocity
components 50\,km\,s$^{-1}$ apart is currently not possible for the
bar/spiral arm interface in the Milky Way, we want to test whether the
scenario is in principle possible by means of studying a nearby galaxy
with more favorable almost face-on geometry.  Comparable regions in
nearby galaxies can provide critical insight to the nature of Galactic
environments like W43.  Much of our understanding in this context
derives from the study of \citet{kenney1991} who first linked star
formation at the end of the bar in M83 to the molecular gas reservoir
traced by CO.  Their detailed examination of the properties and
kinematics of the molecular gas at the time, although limited to
rather coarse resolution, suggested that the change in orbit structure
from the bar to the spiral may lead to the observed local build-up of
gas, which might then favor a burst of star formation.  Soon
thereafter, \citet{rand1992} found supporting evidence for this
picture, showing that the intense star formation observed at the end
of the bar in M51 is linked to locally high (molecular) gas surface
densities.  Since then, numerous observational and theoretical studies
have dealt with galactic bars and their dynamical importance for star
formation as well as galactic dynamics (e.g.,
\citealt{martin1997,jogee2005,verley2007,athanassoula2013,zhou2015}).

But it is still unclear how frequently star formation of this kind
occurs, which types of bars and spirals may be most conducive, or
whether it can explain environments like W43 in our own Milky Way.  To
develop this picture further requires in-depth study of the gas
kinematics at more such interfaces, which are still poorly constrained
in our Milky Way, as well as in external galaxies.

In this paper, we seek to explore potential bar/spiral arm
interactions in an exemplary external galaxy that allows a better
identification of associated velocity components because of a geometry
that is near to face-on.  As a starting point we used the HERACLES
database that provides sensitive CO(2--1) images of several tens of
nearby galaxies \citep{leroy2009}. Additionally, all HERACLES galaxies
have extensive multi-wavelength coverage from the X-ray to the radio
regime from the SINGS \citep{kennicutt2003} and KINGFISH
\citep{kennicutt2011} efforts. Among the barred galaxies in the
HERACLES sample, only two galaxies show the desired geometry, i.e., a
stellar bar with strong spiral arms emanating from the bar ends and
associated massive star formation. NGC3627 is one of these two
galaxies and has already high quality published molecular gas
observations available \citep{paladino2008,leroy2009}. The well-known
barred spiral galaxy NGC3627 exhibits strong burst signatures at the
two bar/arm interfaces in the north and south as visible for example
in the 70\,$\mu$m Herschel map (Fig.~\ref{heracles} left panel,
\citealt{kennicutt2011}). These active bar/arm interaction zones
resemble the structures found in W43 in our Milky Way
well. Furthermore, the location of NGC3627 is very favorable for our
study: the orientation is almost face-on (inclination angle of 65\,deg
with a position angle $\theta_{PA}$=170$^o$, \citealt{chemin2003}),
the object is relatively nearby at a distance of 11.1 Mpc (e.g.,
\citealt{saha1999}) where spatial resolution elements of $1''$
correspond to linear scales of $\sim$54\,pc, and the systemic velocity
is $V_{sys}$ =744 km s$^{-1}$ \citep{casasola2011}. More details will
be given in section \ref{dynamics}. Although these linear scales are
obviously much larger than what can be achieved within our Milky Way,
the almost face-on nature of this galaxy eases the interpretation of
the data tremendously.

Using near-infrared imaging from the Spitzer Survey for Stellar
Structure (S$^4$G, \citealt{sheth2010}), \citet{buta2015} classified
NGC3627 as a $\rm SB_x(s)b pec$, i.e. a strongly barred galaxy with a
boxy/peanut bulge. NGC3627 has a stellar mass of $\rm
log(M_{\star}(M_{\odot}))\approx10.8$ and a specific star formation
rate ($\rm sSFR=SFR/M_{\star}$) of $\rm
log(sSFR(yr^{-1}))\approx-10.3$, i.e., it lies on the local main
sequence of star forming galaxies \citep{salim2007}.  As we are
interested in a qualitative comparison of signatures for the stellar
bar/spiral arm region, an exact match in properties between NGC3627
and the Milky Way is not necessary.

\begin{figure*}[ht]
\includegraphics[width=0.99\textwidth]{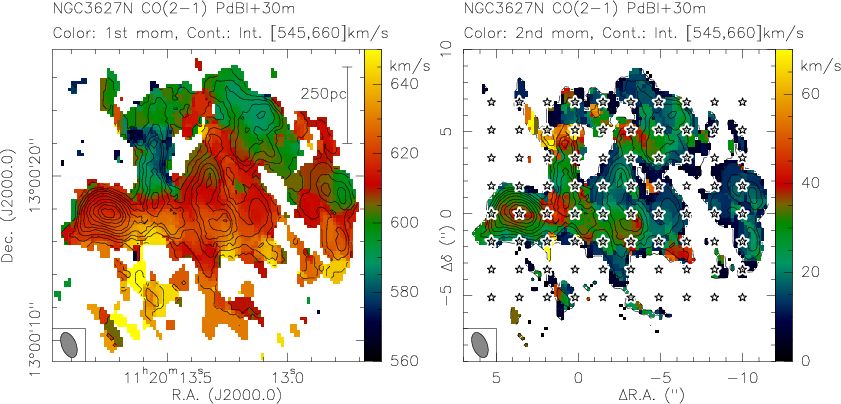}\\
\includegraphics[width=0.99\textwidth]{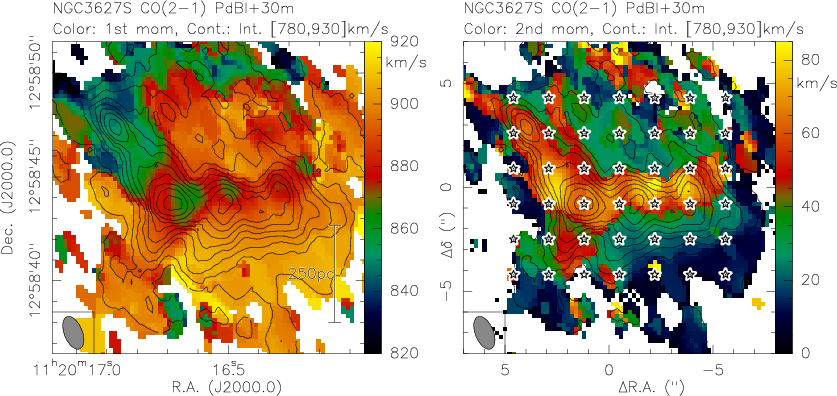}
\caption{Combined PdBI+30m CO(2-1) data toward NGC3627 north and
  south (top and bottom panels, respectively). The colors show in the
  left and right panels the 1st (intensity-weighted peak velocities)
  and 2nd moments (intensity-weighted line widths), respectively. The
  contours present the integrated intensities in the velocity regimes
  indicated above each panel from 5 to 95\% of the peak intensities
  (in 10\% steps). The stars in the right panels show the positions
  for spectra extraction. Example spectra are shown in
  Fig.~\ref{spectra}, and fit results are presented in Tables
  \ref{fits_n} and \ref{fits_s}.}
\label{mom_co21}
\end{figure*}

The two interface regions between the bar and inner spiral arms at the
northern and southern ends of the NGC3627's stellar bar have been
observed with the Plateau de Bure Interferometer (PdBI) in the
CO(1--0) and CO(2--1) emission by \citet{paladino2008}. While they
concentrated on the connection between radio continuum, CO and
8\,$\mu$m emission, we now analyze the kinematics of the gas.  We
re-calibrate and re-image these data (see next section), and in
particular we add the CO(2--1) short spacing information from the
HERACLES data \citep{leroy2009} to minimize flux and imaging
artifacts. Figure \ref{heracles} (right panel) presents the
single-dish HERACLES CO(2--1) data as a first moment map (intensity
weighted velocities) with the integrated intensity contours. This
image shows the bar-spiral structure as well as the rotational
velocities of the gas.  The primary beams of the PdBI observations are
marked as well.

Using the PACS 70$\mu$m image from KINGFISH \citep{kennicutt2011}, we
assess the star formation rate of the NGC3627 southeast bar end
region. To do this, we convolve the PACS image to have a round
Gaussian beam of FWHM $8\arcsec$. At this resolution ($\sim 350$~pc)
the star-forming region is largely unresolved. For the 70$\mu$m-to-SFR
conversion quoted by \citet{kennicutt2012}, the bar end has a SFR of
$\approx 0.23$~M$_\odot$~yr$^{-1}$ inside the $8\arcsec$ Gaussian
beam. This compares relatively well to the $\approx
0.01-0.1$\,M$_{\odot}$yr$^{-1}$ for W43 \citep{nguyen2011} given the
degree of uncertainty in extragalactic star formation rate estimates,
especially for individual regions.

With the goal to better understand the dynamics of bar/arm interfaces
in an external galaxy and set this into context to our Milky Way, we
here study the nearby galaxy NGC3627 as an excellent example case with
a comparable configuration to the Milky Way. This paper is structured
as follows: Section 2 describes the combination of the Plateau de Bure
Interferometer and 30\, single-dish data. Section 3 then outlines the
basic morphological and spectral signatures of the spectral line
data. Section 4 deals with the dynamical interpretation of the
multiple velocity components at the bar/arm interface in the framework
of crossing orbit families corresponding to the bar and arm. Finally,
Section 5 discusses these results as potential reasons for enhanced
star formation activities at bar/arm interfaces, as well as it sets it
into context with our Milky Way and particularly W43. Section 6 then
summarizes our results.

\section{Observations} 
\label{obs}

\begin{figure*}[ht]
\includegraphics[width=0.49\textwidth]{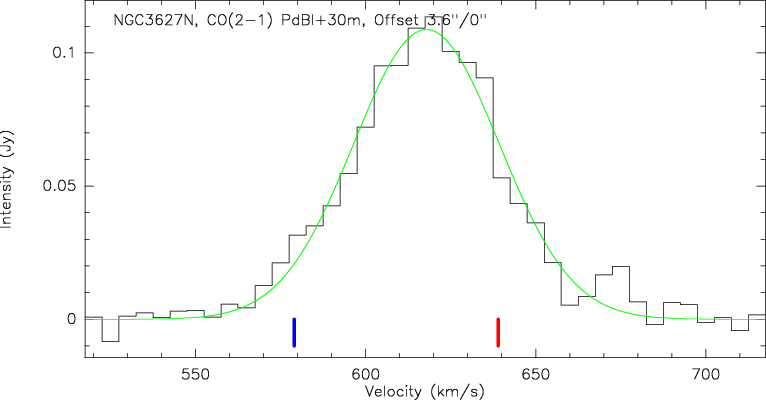}
\includegraphics[width=0.49\textwidth]{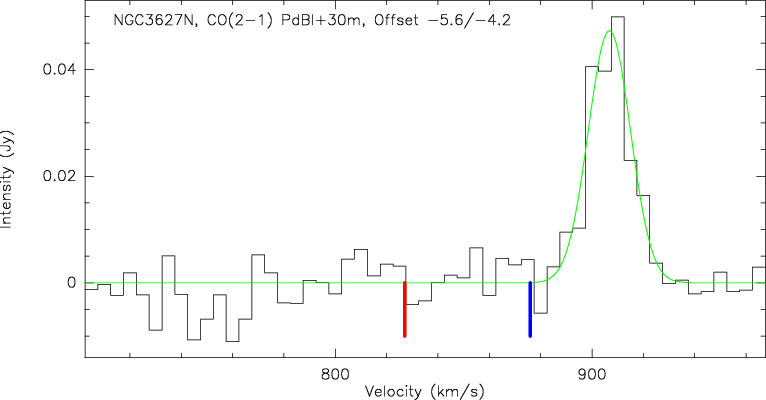}\\
\includegraphics[width=0.49\textwidth]{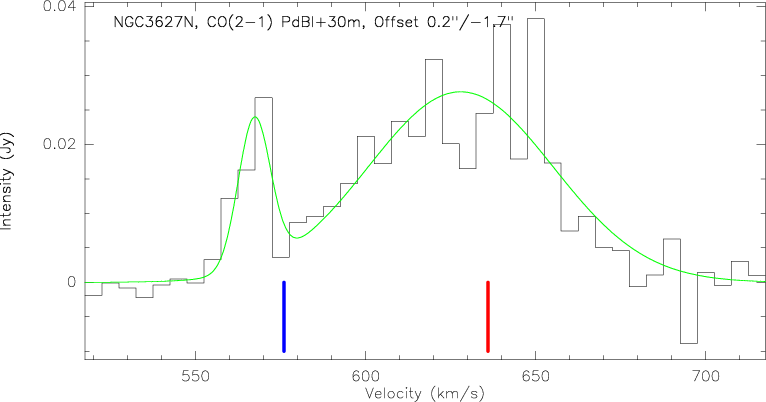}
\includegraphics[width=0.49\textwidth]{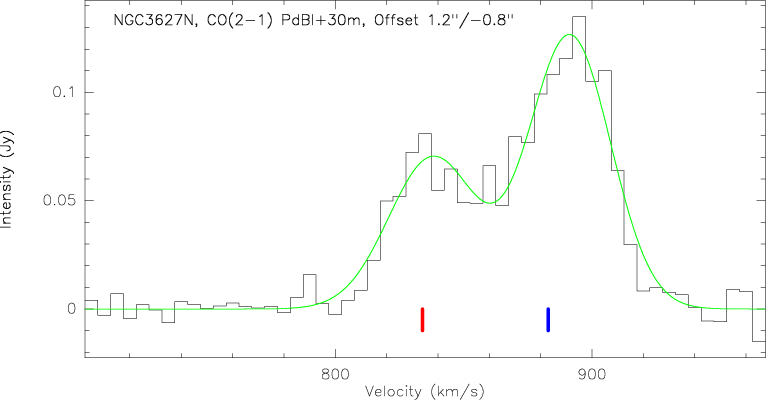}\\
\includegraphics[width=0.49\textwidth]{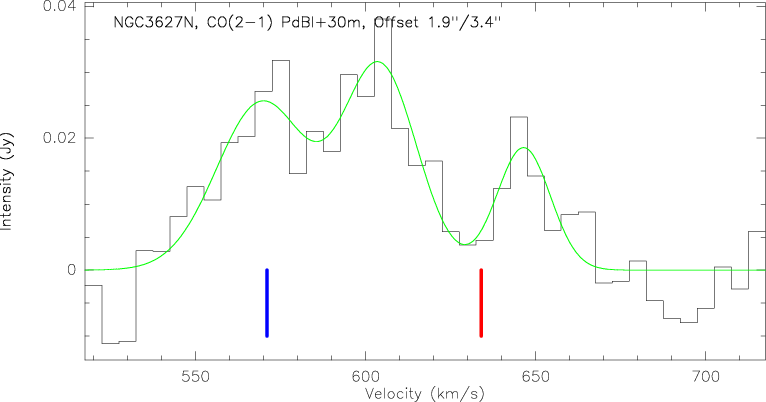}
\includegraphics[width=0.49\textwidth]{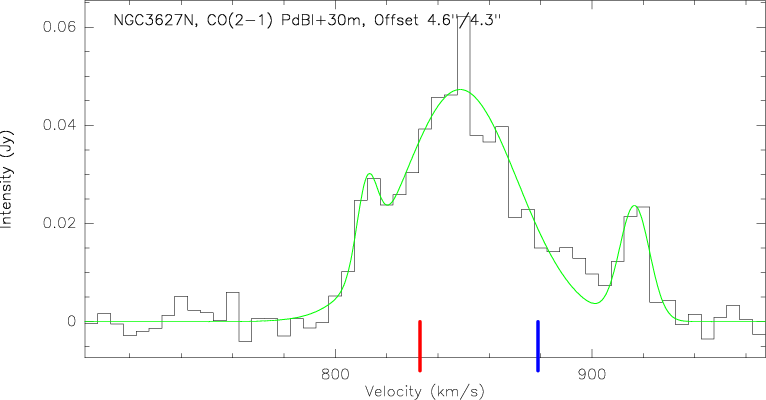}
\caption{PdBI+30m CO(2-1) example spectra toward NGC3627N (left) and
  NGC3627S (right). The offsets of the positions are marked (see also
  Fig.~\ref{mom_co21}). The green lines present the Gaussian fits to
  the data. The corresponding FWHM and peak positions are reported in
  Tables \ref{fits_n} and \ref{fits_s}. The blue and red lines at
    the bottom of each panel mark mean values of the velocities for
    the bar 1 scenario and the spiral described in section
    \ref{dynamics}. The corresponding full velocity ranges are marked
    in Figures \ref{fig:Allprofs} and \ref{fig:NSprofs}.}
\label{spectra}
\end{figure*}

\citet{paladino2008} present the PdBI for the first time, however,
they focus on the CO(1--0) observations and do not complement the
data with short spacing information. The PdBI observations were
carried out on March 20, 2005 with 6 antennas in the C configuration.
At that time, 1.3 and 2.6\,mm data could be taken simultaneously, and
the observations targeted the CO(1--0) and (2--1) lines, respectively.
The two frequencies were centered at 114.997 and 229.88\,GHz which are
the transition rest frequencies at an assumed $v_{\rm{lsr}}$ of
$\sim$712.6\,km\,s$^{-1}$. With four overlapping 160\,MHz spectral
units, the correlator covered 580\,MHz for each line (corresponding to
1512 and 756\,km\,s$^{-1}$ for the (1--0) and (2--1) lines,
respectively). Bandpass and flux calibration were conducted with
1055+018 and MWC349, and the gains were calibrated with regular
observations of 1116+128.

To complement the PdBI CO(2--1) data with the missing short spacing
information, we used the IRAM\,30\,m single-dish observations from the
HERACLES survey \citep{leroy2009}. These data are available for
download at www.mpia.de/HERACLES. The HERACLES CO(2--1) data were
further processed in GILDAS/CLASS and finally combined and imaged with
the PdBI data within the mapping part of the GILDAS package.

In the following, we only use the combined PdBI+30\,m CO(2--1) data.
The phase reference centers for NGC3627N and NGC3727S are
  R.A.(J2000.0) 11:20:13.50, Dec.(J2000.0) 13:00:17.70 and
  R.A.(J2000.0) 11:20:16.60, Dec.(J2000.0) 12:58:44.50. respectively.
The final channel spacing of these is 5\,km\,s$^{-1}$. The
$1\sigma$ rms and synthesized beam values for the merged PdBI+30\,m
CO(2--1) data are $\sim$6\,mJy\,beam$^{-1}$ and $1.66''\times 0.87''$
(P.A. 22\,deg). At the given distance of 11.1\,Mpc, this corresponds
to an approximate linear resolution of $\sim$68\,pc.

\section{Results}

\subsection{Morphological and spectral structures}
\label{structures}

Figure \ref{mom_co21} present the 1st and 2nd moment maps
(intensity-weighted peak velocities and line width) of the combined
PdBI+30\,m CO(2--1) data toward the two target regions at the northern
and southern ends of the galactic bar in NGC3627. In the following, we
refer to the two regions as NGC3627N and NGC3627S. 
Both regions are detected at high significance in the CO(2--1)
emission and the moment maps already reveal a wealth of kinematic
structure. Toward the northern end of the bar, we find an
intensity-weighted peak velocity spread between $\sim$550 and
$\sim$660\,km\,s$^{-1}$, whereas it is even larger in the southern
region between $\sim$760 and $\sim$915\,km\,s$^{-1}$. While we see
velocity gradients across both regions, the 2nd moment maps also show
that most of the broadest lines are found toward the integrated
intensity peaks, hence toward the largest column density positions.

A different way to look at the kinematic properties of the gas is by
inspecting the spectra toward selected points in the fields.
Therefore, we extracted the combined PdBI+30m CO(2--1) spectra toward
regularly spaced grids in NGC3627N and NGC3627S that are separated by
the approximate beam size of $1.7''$ as marked in Figure
\ref{mom_co21}. Figure \ref{spectra} presents example spectra and
Gaussian fits to them, the fit results are listed in Tables
\ref{fits_n} and \ref{fits_s}. To fit the Gaussians, we averaged
spectra over a slightly larger area than the beam size. We included
all spectra with offsets $\pm 0.5''$ from the respective position
which then results in a fitted area corresponding to a $\sim$1.6$''$
beam size, a factor of 1.625 larger in area than for the nominal
resolution. This reduces the average rms in the fitted spectra to
$\sim$4.5\,mJy\,beam$^{-1}$.  The number of Gaussian components to be
fitted was visually identified individually for each spectrum. In most
cases, the selection of number of components was unambiguous and
straightforward (Figure \ref{spectra}), but there exist also positions
where one could find other solutions: for example, the spectrum at
position $0.2''/-1.7''$ in NGC3627N may indicate more than 2
components (Fig.~\ref{spectra}, 2nd panel to the left), however, the
Gaussian fitting algorithm did not converge with more than 2
components. Therefore, we restricted ourselves to the lowest number of
reasonable components in such cases. While very broad components
(Tables \ref{fits_n} and \ref{fits_s}) may in reality consist of more
components, for our interpretation and analysis that is less important
because we are mainly interested in the dominating components that may
stem from the bar and/or spiral. Multiple and blended components are
also discussed in section \ref{comparison}.

The only adjusted parameter for each fit is the number of Gaussian
components, all other parameters, in particular the peak velocities
$v_{\rm{peak}}$ and the full-width-half-maximum line widths $\Delta v$
for each component are free parameters. In general, we fitted
  Gaussians only to peak flux densities above the $3\sigma$ level of
  13.5\,mJy\,beam$^{-1}$. In a small number of cases ($\sim 8$\% out
  of 207 features in total) we allowed also slightly lower peak flux
  densities with the lowest value at 9.2\,mJy\,beam. However, in all
  these cases, visual inspection of the spectra confirmed the reality
  of the features because they are existing over several contiguous
  channels. The strongest peaks exhibit peak flux densities
  $>100$\,mJy\,beam$^{-1}$ (maximum 142\,mJy\,beam$^{-1}$), and we can
  detect peak-to-peak flux density differences between neighboring
  features of around a factor 10. The smallest identified separation
  between adjacent peaks is $\sim$20\,km\,s$^{-1}$ (Tables
  \ref{fits_n} and \ref{fits_s}).

We find a variety of features: Some spectra are single-peaked but many
also exhibit clearly spectrally resolved multiple peaks towards
individual positions. The derived full-width-half-maximum (FWHM or
$\Delta v$) values for the different spectra also cover a relatively
broad range of FWHM values between $\sim 5$\,km\,s$^{-1}$ at the
narrow end (which is also our spectral resolution limit), and values
up to 100\,km\,s$^{-1}$. While the broadest lines likely consist of
multiple components that are difficult to separate, the narrow end
with several values between 10 and 20\,km\,s$^{-1}$ is approximately
in agreement with what would be expected from Larson's line-width
$\Delta v$ size $L$ relation $\Delta v \sim 2.355\times 1{\rm
  km\,s}^{-1}\left( \frac{L}{1\rm{pc}}\right)^{0.5}$, where $\Delta v$
is the velocity FWHM and $L$ the root-mean-square size
\citep{solomon1987}.  Using the average synthesized beam FWHM of
$1.3''$ as an estimate for the size $L\sim 1.3''/2.355$, this results
in a linear $L$ of $\sim 70/2.355$\,pc or an approximate $\delta v$ of
12.8\,km\,s$^{-1}$.  Taking into account that the size estimated this
way is most likely a lower limit since the peak regions are no point
sources but extended (see Fig.~\ref{mom_co21}), we conclude that the
narrower lines observed in NGC3627 are broadly consistent with
Larson's relation and hence resemble typical molecular clouds.

\section{Dynamical interpretation}
\label{dynamics}

\begin{figure}[ht]
\begin{center}
\includegraphics[width=0.49\textwidth]{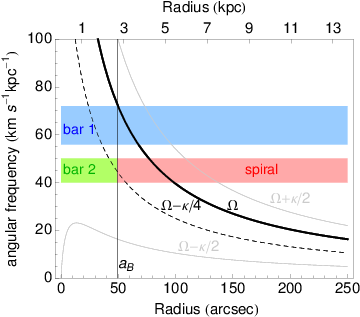}
\end{center}
\caption{Angular frequency curves in NGC 3627: $\Omega$ (thick black),
  $\Omega\pm\kappa/2$ (light gray), and $\Omega-\kappa/4$ (dashed).
  The vertical line marks the end of the bar \citep{chemin2009} with
  length $a_B=49''$. The red horizontal box shows the range in spiral
  speeds as measured from gas kinematics (see text).  The blue
  horizontal box indicates the range of bar speeds in scenario 1,
  suggested by \citet{chemin2009} while the green box shows the bar
  speed in scenario 2 (same speed as the spiral). The different
    bar and spiral scenarios are labeled in the plot.}
\label{fig:freq}
\end{figure}

In this section we examine whether the observed velocity components
are consistent with expected streaming motions in the presence of the
stellar bar and spiral arms. Potential implications for enhanced star
formation activity and the connection to our Milky Way will be drawn
in Section \ref{discussion}.

At each position $(x,y)$ relative to the galaxy center, we assume that
the centroid of the line profile reflects both radial and azimuthal
components $v_r$ and $v_\phi$ in projection, i.e.
\begin{equation}
V_{los}=V_{sys}+\left[v_r\sin{(\theta-\theta_{PA})}+v_\phi \cos{(\theta-\theta_{PA})}\right]\sin{i} \label{eq:los}
\end{equation}
where $\theta$=$\arctan{(y/x)}$ measures the offset from the kinematic
major axis position angle $\theta_{PA}$ (by convention measured in the
anti-clockwise direction with respect to the major axis of the
receding half of the galaxy) and $v_\phi$ includes the circular
velocity $V_c$.  We adopt the inclination $i$=65$^o$,
$\theta_{PA}$=170$^o$ and the kinematic center as determined by
\citet{chemin2003} and take the systemic velocity $V_{sys}$ =744 km
s$^{-1}$ following \citet{casasola2011}.


Our goal in what follows is to develop a model for $v_r$ and $v_\phi$
at the end of the bar that we can compare with the observations.  In
this particular galaxy, given the alignment of the bar morphological
major axis with the disk kinematic major axis (so that
$\theta$$\sim$$\theta_{PA}$), the comparison of a model for $v_r$ and
$v_\phi$ at the end of the bar with observations is considerably
simplified.  Typically, this orientation can inhibit the extraction of
kinematic information related to the bar strength and pattern speed
(i.e., \citealt{rand2004}), since characteristic strong radial motions
directed along x1 orbits in the bar potential are directed almost
entirely out of the line of sight.  But for studying the bar end, this
is in fact ideal: Here the orbits of stars and gas are characterized
by primarily azimuthal motions (and negligible radial motions).  Thus,
in this favorable set-up, the line-of-sight velocity captures almost
entirely the full picture of azimuthal motions in the bar potential.

The dominant contribution to the line-of-sight velocity originates
from circular motion about the galaxy center, projected on the plane
of the sky.  We adopt the rotation curve model of \citet{chemin2003}
that extends out to $\sim$90", beyond which it matches closely the
circular velocities modeled by \citet{trachternach2008}.  Figure
\ref{fig:freq} shows the angular frequency curves $\Omega$=$V_c/R$,
$\Omega\pm\kappa/2$ and $\Omega-\kappa/4$, where the radial orbital
oscillation frequency $\kappa$ is defined by
\begin{equation}
\kappa^2=\left(R\frac{d\Omega^2}{dR}+4\Omega^2\right).
\end{equation}
These curves are shown together with two models for the relation
between the bar and spiral pattern described in more detail below.
These two models correspond to different azimuthal streaming near the
end of the bar, which we assume to additionally contribute to the
observed line-of-sight velocity.

\subsection{The combined bar and spiral system}

Depending on position, the line-of-sight velocity profiles near the
bar end will contain information related to azimuthal motions directed
along orbits in the presence of either the bar or spiral, or both.
When gas populates intersecting bar and spiral orbits, two velocity
components will emerge (although they may not be spatially
resolvable).  The locations of the individual bar and spiral
components, and their relation to each other, depend on the relation
between the bar and spiral pattern speeds.  We consider two
possibilities, one in which the bar and spiral rotate with the same
speed, and a second, in which the bar and spiral rotate with distinct
speeds (although they may be dynamically coupled).  As described
below, the latter corresponds to a scenario in which the bar extends out
to its corotation radius $R_c$, the location where the bar pattern
speed $\Omega_p$ is equal to the disk angular rotation $\Omega$.  In
the second scenario, the bar ends well inside corotation.

Constraints on the likelihood of one or the other scenario in this
particular galaxy are varied, mostly as a result of the orientation of
the bar, which leads to ambiguous results with kinematics-based
techniques.  The direct, model-independent Tremaine-Weinberg method
(TW; \citealt{tremaine1984}) uses departures from axisymmetry traced
by a continuity-obeying kinematic tracer (e.g., HI or CO) along the
line-of-sight to measure the pattern speed, and thus provides little
constraint on the speed of the bar. The pattern speed
$\Omega_{P,s}$=50 km\,s$^{-1}$ kpc$^{-1}$ measured by
\citet{rand2004}, though, provides us with an estimate for the spiral
speed.

Although we cannot rule out that the bar drives a spiral with the same
speed based on the TW method, we have other reason to expect that the
bar may rotate with a much higher speed.  Stellar orbit theory
predicts that bars end at or near their corotation radii (e.g.,
\citealt{contopoulos1980}), which is the location where the pattern
speed $\Omega_p$ is equal to the disk angular rotation $\Omega$.  We
can estimate the speed that would be necessary in order for the bar
length $a_B$ to fall within the predicted range, i.e.,
1$<$$a_B/R_c$$<$1.4, by inspection of the angular rotation curve
$\Omega$ as demonstrated in Figure \ref{fig:freq}.

\citet{chemin2003} previously suggested a pattern speed $\Omega_{P,B}$
in the range 56-72 km\,s$^{-1}$ kpc$^{-1}$, based on the requirement
that 1$<$$a_B/R_c$$<$1.4.  Corotation at this location in the disk was
later confirmed from an analysis of gravitational torques by
\citet{casasola2011}.  Meanwhile the spiral appears to rotate at a
lower speed, likely in the range 40$<$$\Omega_{P,S}$$<$50
km\,s$^{-1}$, according to the results of the TW method
\citep{rand2004} and the harmonic decomposition of the HI velocity
field by \citet{trachternach2008}, which clearly shows the signature
of corotation at $\sim$100''.

This set of bar and spiral speeds are consistent with
the mode-coupling scenario \citep{masset1997}, in which the overlap of
two patterns at resonance leads to the efficient transfer of energy
and angular momentum between them.  Figure \ref{fig:freq} shows that
the bar corotation (where the blue lines intersect with the black
curve, at the bar end) overlaps with the inner 4:1 resonance of the
spiral (where the spiral pattern speed $\Omega_p$=$\Omega-\kappa/4$,
i.e., the red band intersects the dark grey curve), which is an
arrangement that has been found previously in nearby observed galaxies 
(e.g., \citealt{meidt2009,rautiainen2005}).  

\subsubsection{Bar streaming motions}
\label{sec:bar}

 Whether or not the bar extends to its corotation radius has
 implications for the line-of-sight velocity expected for gas motions
 directed along orbits supporting the bar.

\underline{Bar 1 scenario: Bar ends at corotation}\\
In the case that the bar extends to its corotation, we can describe
streaming using the equations of motion specifically at the corotation
resonance of a weak bar perturbation (i.e., \citealt{binney1987},
eq.~3-123).  These imply that the radial and azimuthal streaming
motions are (nearly) identically zero and the azimuthal component of
the velocity equals the circular velocity in the disk\footnote{In the
  case of libration at the bar end (rather than circulation about the
  galaxy center; i.e., \citealt{binney1987}), the trapped stars or gas
  still overall rotate with the bar.  Thus, even with the (small)
  additional component $\Omega \epsilon$ for a libration of size
  $\epsilon$ eq. (\ref{eq:bar1}) remains a good approximation.}, i.e.
\begin{equation}
v_\phi^{B,1}=V_c.  \label{eq:bar1}
\end{equation}

\begin{figure}[htb]
\includegraphics[width=.99\linewidth]{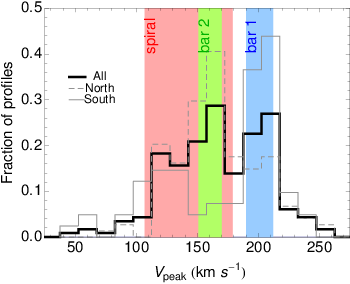}
\caption{Histogram showing the relative frequency of the centroid
  velocities of all fitted velocity components (PdBI+30m CO(2-1)
  data).  Velocities measured in the north and south are shown in gray
  dashed and solid lines, respectively, whereas the solid black line
  takes all components into consideration. The centroid velocity is
  shown transformed to the galaxy-plane (eq. \ref{eq:los}). Overlaid
  rectangles represent the different predictions of the models
  described in section 4. Two predictions for the bar are shown: the
  blue rectangle shows $V_\phi^{B,1}$ in eq.~3 from scenario 1, in
  which the bar has a higher pattern speed than the spiral
  $\Omega_b$$>$$\Omega_s$, while the green rectangle shows
  $V_\phi^{B,2}$ in eq.~6 from scenario 2, in which the bar and spiral
  have the same pattern speed $\Omega_b=\Omega_s$. The red rectangle
  shows the prediction for the spiral motions corresponding to
  $V_\phi^S$ in eq.~8.  Velocities spanned by the spiral illustrate a
  range of spiral arm strengths corresponding to
  $\Sigma_a/\Sigma_0$=0.5-1.5.}
\label{fig:Allprofs}
\end{figure}

\begin{figure*}[htb]
\includegraphics[width=.49\linewidth]{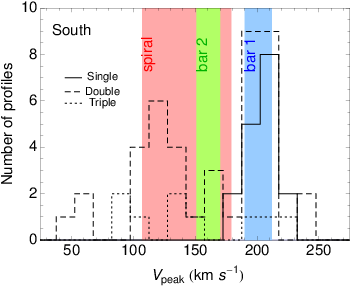}
\includegraphics[width=.49\linewidth]{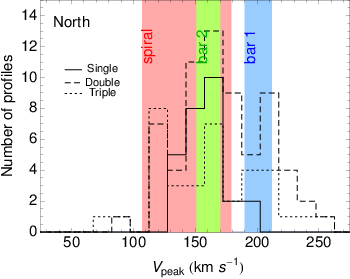}
\caption{Histograms of the centroid velocity of each fitted component
  extracted in the north (right) and south (left). The line style
  indicates the number of velocity components fitted along each
  line-of-sight: single (solid), double (dashed) and triple or more
  (dotted).  The centroid velocity is shown transformed to the
  galaxy-plane as in Figure \ref{fig:Allprofs}.  Overlaid colored
  rectangles show the predictions of the model for the spiral, the bar
  in scenario 1 and the bar in scenario 2 as in Figure
  \ref{fig:Allprofs}.}
\label{fig:NSprofs}
\end{figure*}

\noindent \underline{Bar 2 scenario: Bar ends inside corotation}\\
In the alternate scenario, in which the bar ends well inside its
corotation radius, we can gain insight into gas flow characteristics
at the bar end by considering the case of a `weak' bar (e.g.,
\citealt{sellwood2001}).  Although the bar in NGC3627 is more
accurately a `strong' bar, motions at the bar end are qualitatively
similar to those in the weak case.

We follow \citet{sellwood2010} to estimate the magnitude of azimuthal
streaming adopting the (weak) bar perturbation to the gravitational
potential
\begin{equation}
\Phi_B=-V_c^2\frac{1-q_\Phi^2}{4q_\Phi^2}
\end{equation}
where $q_\Phi$ is the axial ratio of the bar potential, which is related to the axial ratio of the density distribution as
\begin{equation}
q\sim 1-3(1-q_\phi)  
\end{equation}
\citep{binney1987}.  We estimate $q$ as $(1-\epsilon_{\rm{bar}})$
using the bar ellipticity $\epsilon_{\rm{bar}}$=0.69 measured from the
S$^4$G 3.6\,$\mu$m image tracing the stellar mass
\citep{cisternas2013}.

In this case, the azimuthal component of the velocity away from
corotation (i.e., \citealt{sellwood2010}) is reduced to
\begin{equation}
v_\phi^{B,2}\sim\left(1-\frac{1-q_\phi^2}{4q_\phi^2}\right) V_C \sim 0.8 V_c.  
\end{equation}
at the bar end.  Although there are non-negligible radial motions
\begin{equation}
v_r^{B,2}\sim\frac{2}{3}\left(1-\frac{1-q_\phi^2}{4q_\phi^2}\right) V_C \sim 0.5 V_c
\end{equation}
they go to zero at azimuths aligned with the bar major axis.  

\subsubsection{Spiral streaming motions}
\label{sec:spiral}

Spiral density wave theory provides the basis for estimating the
magnitude of spiral streaming motions, which we assume are the same in
the two scenarios (since the spiral pattern speed is the same in
both).  Following \citet{binney1987} in the tight-winding
approximation the azimuthal velocity component associated with the
spiral is
\begin{equation}
v_\phi^S=V_c\left(1-\frac{\sin{m(\phi-\phi_s)}}{m\cot{i_p}}\frac{\Sigma_a}{\Sigma_0}\right)  \label{eq:spiralphi}
\end{equation} 
and the radial component is
\begin{equation}
v_r^S=\frac{-m R (\Omega-\Omega_p)\cos{m(\phi-\phi_s)}}{m\cot{i_p}}\frac{\Sigma_a}{\Sigma_0}
\label{eq:spiralr}
\end{equation} 
where the circular velocity is $V_c$=$R\Omega$, $\Omega_p$ is the
spiral pattern speed, $i_p$=34$^o$ is the spiral arm pitch angle
\citep{lavigne2006}, $m$=2 since there are two spiral arms, and
$\Sigma_a/\Sigma_0$ is the ratio of the perturbed to unperturbed
surface mass density.  We estimate $\Sigma_a/\Sigma_0$ using the
arm-interarm contrast, which we expect to fall between 2-3 (so that
$\Sigma_a/\Sigma_0$$\sim$1-2) given this galaxy's SABb classification,
using the 3.6\,$\mu$m contrast for this type measured by
\citet{elmegreen2011} and assuming a constant 3.6 $\mu m$
mass-to-light ratio (e.g., \citealt{meidt2014})\footnote{We have
  confirmed that this is consistent with the arm contrast measurable
  directly from the stellar mass map for this galaxy derived by
  \citet{querejeta2014} from its S$^4$G Spitzer/IRAC 3.6\,$\mu$m
  image.}.

Given the alignment (zero phase offset) between the end of the bar and
the start of the spiral arms, we consider motions near the spiral arm
potential minimum where $\phi\approx\phi_s$.  According to eq.
(\ref{eq:spiralphi}) azimuthal streaming motions should go to zero.
However, the gas response can be complicated, with shocking and
viscous and gravitational torques that introduce non-zero azimuthal
motions in close proximity to the spiral minimum, such as observed in
M51 \citep{shetty2007}.  Models of gas streaming in the presence of a
density wave qualitatively predict that $v_\phi$ varies rapidly as gas
encounters the arm, passing through zero on switching from negative
(in the interarm) to positive immediately upon exiting the arm (e.g.,
\citealt{roberts1987,shetty2007}). In the slightly upstream position
at which the observed line profiles are extracted (see Figures
\ref{mom_co21} \& \ref{heracles}), we expect a maximum of
$v_\phi^S$$\sim$ 0.5$V_c$= and $v_r^S$$\sim$ -25 km\,s$^{-1}$.

\begin{figure*}[htb]
\includegraphics[angle=-90,width=.99\linewidth]{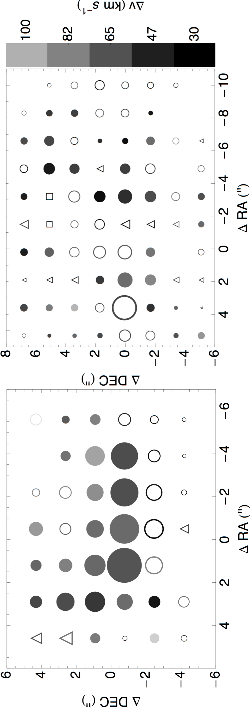}
\caption{Map of the locations of the different velocity profiles in
  the south (left) and north (right) where spectra are fitted with
  single (open circle), double (filled circle), triple (triangle) or
  more (square) components (see for comparison Fig.~\ref{mom_co21}).
  The size of the symbol is scaled to the integrated CO intensity
  while the gray scaling indicates the effective velocity width of the
  profile shown by the scale-bar to the right (estimated by summing
  the $\Delta V$ measured for each individual component in
  quadrature). The same scaling is adopted for the northern and
  southern bar ends.}
\label{fig:NSmaps}
\end{figure*}

\begin{figure*}[t]
\includegraphics[width=.49\linewidth]{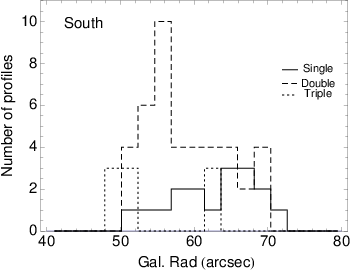}
\includegraphics[width=.49\linewidth]{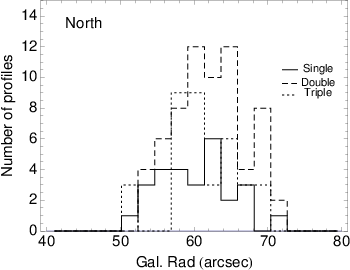}\\
\includegraphics[width=.49\linewidth]{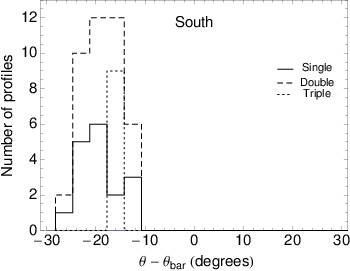}
\includegraphics[width=.49\linewidth]{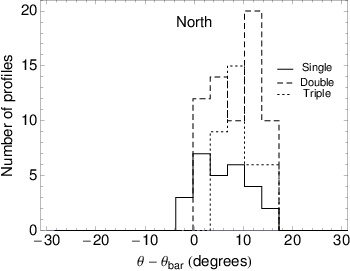}\\
\includegraphics[width=.49\linewidth]{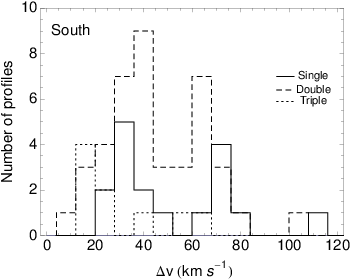}
\includegraphics[width=.49\linewidth]{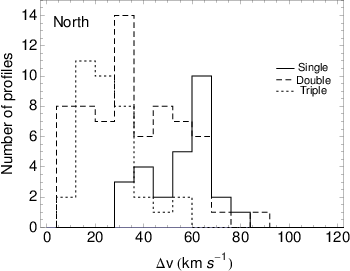}
\caption{Histograms of galactocentric radius (top) and azimuthal
  angle (relative to the bar major axis; middle) at which spectra are
  extracted towards the north (right) and south (left) ends of the bar
  in NGC3627. The bar end is located at $a_B=49''$ according to
  \citet{chemin2009}. The line style indicates the number of velocity
  components fitted along each line-of-sight: single (solid), double
  (dashed) and triple or more (dotted). Histograms of the FWHM $\Delta
  v$ of each fitted component are shown in the bottom
  row.}
\label{fig:NSstats}
\end{figure*}

Note that radial streaming motions would be considerably larger for a
lower spiral pattern speed, whereas azimuthal motions would be
unchanged.  As we are most sensitive to the latter along the line of
sight with this particular set of disk and bar orientations, we cannot
distinguish the proposed scenario, in which the spiral pattern speed
is $\Omega_p \sim$ 50\,km\,s$^{-1}$ kpc$^{-1}$, from one in which it
is closer to $\Omega_p \sim$ 25\,km\,s$^{-1}$ kpc$^{-1}$.  Such a low
value has been suggested by \citet{reuter1996}, taking the corotation
$R_c=216''$ identified with the \citet{canzian1993} method and is also
consistent with the value measured by \citet{rand2004} specifically
for the southern extension of the western arm, which appears to lie
off the plane of the rest of the disk.  We will not consider streaming
motions predicted in the case of this low speed further given they are
likely applicable only at larger galactocentric radius and would
further be identical to those predicted in the azimuthal direction in
our nominal scenario.  

\subsection{Comparison to observations}
\label{comparison}

As argued in the previous section, the line-of-sight velocity at the
bar end in NGC3627 is dominated by azimuthal motions, and we expect
two distinct kinematic components when gas populates bar and spiral
orbits.  Thus, in our bar 1 scenario, in which the bar ends
at its corotation radius, we expect one component from the bar to be
centered at $\sim V_{sys}\pm V_c(a_B)\sin{i}$. This component should
exhibit as much as 80 km\,s$^{-1}$ offset from a second component
associated with the spiral, at $\sim V_{sys}\pm0.5V_c(a_B)\sin{i}$. 

Figures \ref{fig:Allprofs} and \ref{fig:NSprofs} show the measured
centroid velocity in comparison to the different bar1, bar 2 and
spiral model intorduced in the previous section.  NGC3627S frequently
shows two velocity components (21 out of 42).  The lower-velocity
component arises from within a range $\sim$100-150\,km\,s$^{-1}$,
whereas the higher-velocity component is more regularly found near an
average of 200$\pm$10\,km\,s$^{-1}$. The single-peak profiles (17 out
of 42) consistently arise near the high 200\,km\,s$^{-1}$ value and
tend to be located offset from the region of brightest CO
emission. Fig.~\ref{fig:NSmaps} visualizes the number of components
with respect to the location, confirming that the single-peaked
profiles avoid the zone of strongest interaction and trace mainly bar
orbits (see right panel of Fig. \ref{fig:NSmaps}). Further statistical
properties of the kinematic gas properties are shown in
\ref{fig:NSstats} and refered to below.

These features are remarkably consistent with the predictions of our
simple model. Based on the higher velocity components in
Fig. \ref{fig:NSprofs}, which are close to the circular velocity
$V_c\approx$200 km s$^{-1}$ at these radii, we can rule out the bar 2
scenario (the green rectangles in Figs. \ref{fig:Allprofs} and
\ref{fig:NSprofs}) in which the maximum velocity does not exceed ~0.8
$V_c$. In contrast to this, we find good agreement between the
observations and our model for velocities at the end of the bar in
scenario 1 (the blue rectangles in Figs. \ref{fig:Allprofs} and
\ref{fig:NSprofs}). The width of the blue and green rectangle
illustrating the predicted bar velocity is set by the range in $V_c$
corresponding to the range in galactocentric radii spaned by the
mapped regions.

The prediction for the spiral (red rectangle in
Figs. \ref{fig:Allprofs} and \ref{fig:NSprofs}), which also agrees
well with the observations, spans a larger range to demonstrate the
effect of variation in the strength of the spiral and/or in the exact
(azimuthal) location at which the profile is extracted relative to the
spiral potential minimum.  Figs.~\ref{fig:Allprofs} and
\ref{fig:NSprofs} illustrate the range of velocities corresponding to
an increase in the strength of the spiral arm from
$\Sigma_a/\Sigma_0$=0.5 to 1.5. This range of spiral strengths yields
azimuthal streaming motions that vary from low
($V_s=33$\,km\,s$^{−1}$) to high ($V_s=$100\,km\,s$^{−1}$). This
  reduces the azimuthal velocity $v_\phi^S=v_c-V_s$ (eq.~8, plotted in
  Figs.~\ref{fig:Allprofs} and \ref{fig:NSprofs}) at these spatial
  locations to $91-172$\,km\,s$^{-1}$ (section 4.1.2.).

  In comparison to the south, spectra in NGC3627N still show
  double-peaked profiles (33 out of 80), but regularly exhibit also
  single components centered near the spiral velocity of
  $\sim$150\,km\,s$^{-1}$ (27 out of 80; Figs. \ref{fig:Allprofs} and
  \ref{fig:NSprofs}). Although the bar 2 model could be consistent
  with the lower grouping of velocities near 150\,km\,s$^{-1}$
  (Fig.~\ref{fig:NSprofs}), measurements in NGC3627S make this
  scenario unlikely. Thus, we associate the lower velocity component
  in the north with the spiral. The lower velocity compoenent is more
  frequent than the higher velocity component around 200 km s$^{-1}$
  (expected for the bar, Fig. \ref{fig:Allprofs}).  In contrast to the
  south, where single peaks are mostly associated with the higher
  velocity bar component, in the north many single-peaked profiles are
  centered close to the lower-velocity spiral component. Indeed, all
  profiles in the north lie further beyond the bar end than in the
  south, by an average of $10''\sim$0.5 kpc (top row of
  Fig. \ref{fig:NSstats}). They may also arise from a different
  location in the spiral potential than the profiles extracted in the
  south, which are found on average 10 degrees further downstream
  (middle row of Fig.~\ref{fig:NSstats}).  According to the model,
  qualitatively in both circumstances the magnitude of the spiral
  streaming motions might be reduced compared to the south where
  profiles appear to arise consistently throughout the bar-spiral
  interaction zone.  This could explain why the lower of the two
  primary velocities observed in the north is nearer to the circular
  velocity, implying smaller spiral streaming motions.

  The northern bar end probed by the observations might also be
  influenced by interaction with the companion galaxy NGC3628. This
  galaxy has an evident impact on the morphology and kinematics of the
  spiral emerging from the northern bar end at large galactocentric
  radius ($R=90''$, \citealt{chemin2003}). These signs of interaction
  suggest that the spiral arm dynamics are more complex, even at
  galactocentric radii $R\sim 50''$, and deviate from our simple
  model.

The number and the widths of velocity peaks along each line-of-sight
contain additional information to support the idea of interacting
clouds on overlapping bar and spiral orbits. For example, at either
bar end a small fraction of spectra shows more than two velocity peaks
(3 out of 42 in the south and 14 out of 80 in the north). On average,
these are narrower ($\Delta v\sim 20$\,km\,s$^{−1}$,
Fig.~\ref{fig:NSstats} bottom row) than either of the two primary
components characteristic of the double- or single-peaked profiles
(typical $\Delta v \sim 40-80$\,km\,s$^{−1}$) and closer to the value
expected from the size-linewidth relation (Sec. 3.1). In all cases, at
least two of the velocities are consistent with the bar-spiral model
while other additional components fall near one or the other of the
two primary velocities, and thus appears to belong to either the bar
or spiral components.

\section{Discussion}
\label{discussion}

\subsection{Enhanced star formation at bar-spiral interfaces?}

In the previous section, we presented a simple model of intersecting
gas-populated orbits supporting the bar and spiral arms that can
explain the double-peaked line of sight velocity profiles observed in
CO(2--1) emission at the end of the bar in NGC3627.  These two
velocity components most regularly appear at the southern bar end.
Kinematics to the north, on the other hand, appear to deviate from the
simple model expectation, due, e.g., either to the influence of the
interaction evident at larger radii or to a genuine underpopulation of
one set of orbits.  Regardless of the nature of gas motions in the
north, it is clear that, over a comparably sized areas at the end of
the bar, the frequency of two distinct components is higher in the
south than in the north.  At the same time, the south shows enhanced
rates of star formation per unit gas mass relative to the north (e.g.,
\citealt{paladino2008}).  It therefore appears that the existence of
double-peaked profiles is connected to enhanced star formation,
presumably through an increase in collisions within the gas.

Here we discuss the mechanisms by which our model of gas motions at
the intersection of the bar and spiral arms can lead to intense star
formation, focussing only on the role of streaming motions in
increasing the likelihood of collisions in the region.  As first
argued by \citet{kenney1991}, orbit crossing between orbit families
alone would seem to be insufficient for building high gas densities
and triggering star formation. The region of potential crossing within
the gas is spatially extended and unlikely to explain the highly
localized bursts of star formation observed.  Instead, collisions
could be enhanced with the help of streaming motions which
specifically transport the gas into the orbit-crossing region.

\begin{table}
\caption{Streaming motions at the bar-spiral interface in NGC 3627}
\begin{tabular}{c|cc|cc}
\hline 
\hline
& \multicolumn{2}{c}{bar} & \multicolumn{2}{c}{spiral}  \\
& $v_\phi$ & $v_r$& $v_\phi$ & $v_r$ \\
\hline
scenario 1: $a_B\approx R_c$& $V_c$ & 0 & 0.5 $V_c$ & $\sim -25 km\,s^{-1}$ \\
scenario 2: $a_B<R_c$& 0.8 $V_c$ & 0& 0.5 $V_c$ & $\sim -25 km\,s^{-1}$ \\
\hline 
\hline
\end{tabular}
\begin{tabular}{l} 
All values are approximate (see text). \\
In scenario 1 (scenario 2) $\Omega_{p,B}\neq\Omega_{p,S}$ ($\Omega_{p,B}=\Omega_{p,S}$).
\end {tabular}
\label{tab:preds}
\end{table}

Table \ref{tab:preds} summarizes the velocities expected in the two
different scenarios considered so far.  As discussed in the previous
section, our comparison rules out the second scenario, which predicts
much lower velocities at the bar end in NGC 3627 than exhibited by the
observed profiles.  But we include it here given that it is thought to
apply in the prototype case of M83.  (\citealt{kenney1991} argued that
the bar corotation radius $R_{CR}$ in M83 lies well beyond the bar
end, and thus the bar and spiral patterns have the same angular speed
at the bar-spiral interface.)

In both scenarios, radial streaming is expected to be minimal at the
interface, $\sim$25\,km\,s$^{-1}$ due to motion associated with the
spiral.  The primary difference between scenario 1 and 2 is the
relative azimuthal velocity between the bar and spiral.  In the first
scenario, the bar and spiral exhibit a 80 km\,s$^{-1}$ difference,
whereas we expect the bar and spiral to have very little relative
azimuthal velocity at the interface in scenario 2.  This is
qualitatively different from what was described by \citet{kenney1991}
at the end of the bar in M83, where it is suggested that streaming is
observed to enhance collision in exactly scenario 2.

The difference arises with the interpretation of the observations.
With only $\sim$400 pc resolution, the interferometric CO(1-0)
observations analyzed by \citet{kenney1991} had insufficient
resolution to reveal multiple velocity components along the line of
sight.
Thus, to estimate the angular velocities at the interface, they
measured single, isolated components that sampled the bar and spiral
individually from neighboring positions.
However, according to the previous section, we do not expect the
angular velocities exhibited by gas populating either set of bar or
spiral orbits to necessarily persist at the actual orbit interface.
For bars that end at corotation, bar streaming will be of order
-$V_c\epsilon$ at small distance $\epsilon$ from corotation but then
go to zero at the interface with the spiral.  So although
\cite{kenney1991} infer a velocity difference in the gas populating
bar and spiral orbits at the interface in their presumed scenario 2,
we suspect that this may not be evident upon closer inspection.  (The
bar-spiral interface in M83 may even be found to better resemble
scenario 1, if the bar and spiral exhibit genuinely different speeds,
unlike assumed by \citealt{kenney1991}.)

So, in contrast to the conclusion of \cite{kenney1991}, we argue that
the likelihood of collision is actually quite low in scenario 2, when
the bar and spiral rotate with the same pattern speed.  Rather, the
large azimuthal velocity difference characteristic of scenario 1, when
the bar and spiral have different speeds, likely enhances collision.
Note that even if the spiral arms were to rotate with the same high
pattern speed as the bar, the likelihood of collision in such a
scenario would still be reduced.  In this case, the spiral arms would
sit entirely outside corotation, thus leading to azimuthal streaming
motions directed in the opposite sense to that described above (even
as they approach zero near corotation).

Numerical simulations by \citet{renaud2013,renaud2015} and
\citet{emsellem2015} model a Milky-Way-like galaxy at sub-pc
resolution to study various aspects of ISM interactions from large
spiral- and bar-scales down to the fueling of central black
holes. Recently, \citet{renaud2015} investigated the impact of bars to
trigger and/or regulate star formation. In this study, they also found
that the leading edges of bars are favorable for converging gas flows
and large-scale shocks (see also \citealt{athanassoula1992}). While
the gas circulates fast along the bar, it slows down and accumulates
at both bar ends. Hence, orbital crowding at the edge of the bar can
then lead to cloud-cloud interactions.

Shear motions may act destructively on the star formation
process. However, in the case of NGC3627, gas surface densities at the
bar end exceed the critical density for stabilization from shear or
Coriolis forces. We have estimated the shear at the bar end adopting
the rotation curve in Figure \ref{fig:freq}. Only background shear due
to differential rotation of the disk material is important at the end
of the bar because shear due to gradients in the motions within the
bar potential go to zero at the bar end, where streaming motions are
reduced (e.g., \citealt{athanassoula2013} and as described by the
analytical model here). Following \citet{elmegreen1993} (and see
\citealt{meidt2013}), one can estimate the shear critical surface
density as $\Sigma_{crit,shear}=2.5 \frac{\sigma A}{\pi G}$ where the
Oort parameter $A =1/2(v_c/R - dv_c/dR)$ measures the shear rate and
$\sigma$ is the gas velocity dispersion.  At the end of the bar,
$R=49''$, we find the shear critical surface density to be
85\,M$_{\odot}$\,pc$^{-2}$.  We compare this to the observed surface
densities in the two mapped regions: The average integrated
intensities of the maps in Figure \ref{mom_co21} are
1.88\,Jy\,km\,s$^{-1}$ and 2.44\,Jy\,km\,s$^{-1}$ for NGC3727N and
NGC3627S, respectively. Assuming Rayleigh-Jeans, this converts to
25.6\,K\,km\,s$^{-1}$ and 33.2\,K\,km\,s$^{-1}$ for the northern and
southern region.  This is equivalent to 113 and
146\,M$_{\odot}$\,pc$^{-2}$, adopting a standard Galactic CO-to-H$_2$
conversion factor (e.g., \citet{hughes2013}), well above the shear
critical density estimated above. Since these are just average values,
the peak surface densities are even higher. Thus, it seems unlikely
that shear at the bar end in NGC3627 is greatly weakening the ability
of gas to form stars.

Based on our findings in NGC 3627, we thus conclude that the strength
of the interaction between gas populating bar and spiral orbits may be
more sensitive to the relative azimuthal speeds of the two components,
rather than the difference in radial motions.  Bursts of star
formation may therefore occur preferentially when the bar and spiral
have two different pattern speeds.

We emphasize that the likelihood of collision is likely further
enhanced given gravitational torquing on the gas (due to the bar and
spiral).  When both the bar and spiral sit inside their own corotation
radii, the torques they exert will lead to continual gas inflow.
Radially inward motions along the spiral supporting orbits might be
particularly conducive to collision.  Note that this would occur in
both scenarios.

\subsection{Comparison with the Galactic bar-arm interaction region
  around W43}

One of the goals of this investigation is to provide a comparison to
Milky Way studies of the bar-arm interaction regions that are hampered
by Galactic projection and line-of-sight contamination problems. As
outlined in the Introduction, the bar-arm interface in the Milky Way
around W43 exhibits several velocity components, a very broad one
between 60 and 120\,km\,s$^{-1}$ consisting of several sub-components,
and a separate one roughly between 30 and 55\,km\,s$^{-1}$. Several
indicators described in section \ref{intro} support a picture of two
physically related clouds at the location of W43, however, this
evidence is only circumstantial and simple projection effects of
clouds in different spiral arms are also capable to explain the
results \citep{nguyen2011,beuther2012a,carlhoff2013,motte2014}.

Several parameters can be compared between the W43 complex and the
clouds in the bar-arm interface of NGC3627. These are the velocity
differences between several cloud components, the line-widths as well
as the peak flux densities of different velocity components. For W43,
\citet{nguyen2011} report a ratio of peak flux densities between the
100 and 50\,km\,s$^{-1}$ components between 3 and 4 (their Figure
2). These are typical ratios we find in NGC3627 as well. Furthermore,
for W43, the velocity difference between both components is
$\sim$50\,km\,s$^{-1}$, and \citet{nguyen2011} report a FWHM of the
$^{13}$CO(1--0) line averaged over an extend of $\sim$170\,pc of
$\sim$22.3\,km\,s$^{-1}$. A direct comparison between the numbers for
W43 and those reported in section \ref{structures} is tricky because
of several reasons: First, the $^{13}$CO(1--0) line in W43 traces
partly different gas components than the $^{12}$CO(2--1) line analysis
presented in the current paper. Furthermore, W43 in our Milky Way is
observed along the line of sight, whereas the NGC3627 clouds are seen
almost face-on. Nevertheless, the measured line widths and velocity
separations approximately agree and are comparable between NGC3627 and
the Milky Way. While cloud-cloud collisions in theoretical simulations
without Galactic bar effects usually take place at lower
velocity-separations ($\leq 20$\,km\,s$^{-1}$, e.g.,
\citealt{dobbs2015}, but see also \citealt{renaud2015} finding
  larger velocity separations including a bar potential), and
\citet{motte2014} concentrate on velocity components in that regime
for their converging gas flow stduy in W43, bar-spiral interface
regions are special in that regard as they pile up much more material
than in other regions of a galaxy. Since velocity separations are
similar in NGC3627 as well as the Milky Way W43 region, even if shear
motions are strong, physical associations of clouds at velocity
differences on the order of 50\,km\,s$^{-1}$ appear
reasonable. Although this does not allow a conclusive answer whether
the two clouds in W43 are indeed physically related or not, the data
in NGC3627 of several clouds with comparable spectral parameters make
the cloud-cloud interface interpretation for W43 at least plausible.

\section{Conclusions}
\label{conclusion}
       
With the aim to gain insight into the nature of the Galactic
bar/spiral arm interface region W43 in our Milky Way, we investigate
as an extragalactic counterpart example region the kinematic
properties of the molecular gas in the bar/interface region of the
nearby, almost face-on galaxy NGC3627. Similar to our Milky Way, we
also find several independent spectral velocity components in a large
number of spectra toward the bar/spiral interface in NGC3627. Modeling
these velocity components as being caused by different orbit families
populating independently the bar and spiral of NGC3627, we find that
solutions with the bar extending to the corotation radius give a
reasonable fit to our data. 

The implications of this are manyfold. The similarity of the spectral
signatures at the Milky Way bar/spiral interface compared with that of
NGC3627 make it plausible thatt he multiple velocity components found
toward W43 may indeed stem from interacting molecular clouds in this
region, and not be located in different spiral arms.  Furthermore, the
extremely active star formation processes at the bar/spiral interface
may exactly be caused by such crossing gas orbits and
interacting/colliding gas clouds. The crossing gas streams may pile up
significant amounts of dense gas which henceforth can collapse and
undergo intense star formation activity. Furthermore, the gas
  surface densities in NGC3627 are so high that shear motions are
  unlikely capable to significantly reduce the star formation
  activity. While this scenario is suggestive, further investigations
of more regions with respect to their gas kinematics as well as star
formation activity are needed to further investigate the importance of
such cloud-collision processes.

\begin{acknowledgements} 
  We like to thank Fabian Walter for feedback and discussion during
  the process of this work. We are also greatful to the HERACLES team
  for making the single-dish data available. Furthermore, we thank the
  referee for carefully commenting on the draft.
\end{acknowledgements}

\bibliographystyle{aa}    

\appendix

\section{Fit results}

\begin{table}[htb]
\caption{Gaussian fit results for NGC3727N}
\label{fits_n}
\begin{tabular}{lrr|lrr}
\hline 
\hline
Offset & $v_{\rm{peak}}$ & $\Delta v$ & Offset & $v_{\rm{peak}}$ & $\Delta v$ \\
$('')$ & (km\,s$^{-1}$) & (km\,s$^{-1}$) & $('')$ & (km\,s$^{-1}$) & (km\,s$^{-1}$) \\
\hline
-10.0/-5.1   & --    & --    & -1.5/0       & 612.5 &  25.3 \\
-10.0/-3.4   & 632.5 &  39.0 &              & 639.9 &  20.3 \\
-10.0/-1.7   & 625.7 &  27.6 &              & 690.3 &  11.3 \\
-10.0/0      & 613.3 &  41.4 & -1.5/1.7     & 611.0 &  32.3 \\
-10.0/1.7    & 603.5 &  28.6 & -1.5/3.4     & 603.7 &  61.1 \\
-10.0/3.4    & 617.4 &  62.3 & -1.5/5.1     & 560.8 &  24.5 \\
-10.0/5.1    & 614.6 &  38.7 &              & 599.4 &  12.8 \\
-10.0/6.8    & --    & --    &              & 623.4 &  20.3 \\
-8.3/-5.1    & --    & --    &              & 651.8 &  12.3 \\
-8.3/-3.4    & --    & --    & -1.5/6.8     & 567.5 &  17.3 \\
-8.3/-1.7    & 594.7 &  18.5 &              & 606.7 &  42.7 \\
             & 628.3 &  27.9 &              & 651.5 &  25.7 \\
-8.3/0       & 608.3 &  48.1 & 0.2/-5.1     & 633.5 &  53.5 \\
-8.3/1.7     & 611.1 &  62.1 & 0.2/-3.4     & --    & --    \\
-8.3/3.4     & 600.0 &  29.7 & 0.2/-1.7     & 567.3 &  11.2 \\
             & 643.0 &  35.3 &              & 627.9 &  64.0 \\
-8.3/5.1     & 591.7 &   6.6 & 0.2/0        & 622.0 &  57.5 \\
             & 619.8 &  43.3 & 0.2/1.7      & 598.6 &  57.0 \\
-8.3/6.8     & 607.6 &  61.6 & 0.2/3.4      & 587.3 &  63.0 \\
-6.6/-5.1    & 582.3 &  23.7 & 0.2/5.1      & 553.2 &  14.8 \\
             & 616.1 &  17.2 &              & 601.6 &  56.3 \\
             & 651.6 &  38.4 & 0.2/6.8      & 567.5 &  20.9 \\
-6.6/-3.4    & 629.1 &  78.2 &              & 602.1 &  28.4 \\
-6.6/-1.7    & 611.2 &  58.4 & 1.9/-5.1     & 582.8 &  19.9 \\
             & 647.9 &  21.6 &              & 607.0 &  11.7 \\
-6.6/0       & 589.6 &   8.1 &              & 640.5 &  33.2 \\
             & 617.8 &  29.9 & 1.9/-3.4     & 570.7 &   7.7 \\
-6.6/1.7     & 565.5 &   9.2 &              & 622.6 &  55.2 \\
             & 609.4 &  38.8 &              & 651.7 &   8.1 \\
-6.6/3.4     & 605.4 &  33.5 & 1.9/-1.7     & 584.4 &  62.7 \\
-6.6/5.1     & 550.8 &   7.9 &              & 642.3 &  33.6 \\
             & 599.0 &  48.6 & 1.9/0        & 604.6 &  54.7 \\
-6.6/6.8     & 589.2 &  48.8 &              & 642.6 &  29.8 \\
             & 609.6 &  12.1 & 1.9/1.7      & 596.3 &  38.2 \\
-4.9/-5.1    & 627.6 &  66.0 &              & 646.2 &  23.1 \\
-4.9/-3.4    & --    & --    & 1.9/3.4      & 569.5 &  31.4 \\ 
-4.9/-1.7    & 625.2 &  47.0 &              & 604.2 &  25.7 \\
-4.9/0       & 575.3 &  35.8 &              & 646.5 &  18.5 \\
             & 618.1 &  28.9 & 1.9/5.1      & 556.0 &  28.7 \\
-4.9/1.7     & 574.8 &  23.8 &              & 602.6 &  26.2 \\
             & 615.3 &  20.4 &              & 647.5 &   9.1 \\
             & 647.8 &  15.9 & 1.9/6.8      & 568.7 &  11.2 \\
-4.9/3.4     & 546.7 &   6.1 &              & 597.4 &  30.7 \\
             & 607.2 &  45.5 &              & 643.7 &  11.9 \\
-4.9/5.1     & 591.4 &  21.2 & 3.6/-5.1     & 591.2 &  19.4 \\
             & 617.6 &  26.4 &              & 611.4 &   8.3 \\
-4.9/6.8     & 590.0 &  25.3 &              & 637.1 &  15.6 \\
-3.2/-5.1    & 574.7 &  23.2 & 3.6/-3.4     & 581.7 &  51.8 \\
             & 627.2 &  46.4 &              & 625.7 &  30.4 \\
-3.2/-3.4    & 619.5 &  70.5 & 3.6/-1.7     & 618.3 &  40.2 \\
-3.2/-1.7    & 546.7 &  12.9 &              & 649.2 &   9.5 \\
             & 624.9 &  49.1 & 3.6/0        & 617.9 &  50.3 \\
-3.2/0       & 537.9 &   5.0 & 3.6/1.7      & 612.9 &  53.4 \\
             & 617.7 &  40.2 & 3.6/3.4      & 577.9 &  20.5 \\
-3.2/1.7     & 612.2 &  28.0 &              & 609.7 &  85.2 \\
             & 645.3 &   8.5 & 3.6/5.1      & 590.3 &  61.9 \\
-3.2/3.4     & 613.7 &  60.1 &              & 647.9 &  27.4 \\
-3.2/5.1     & 528.3 &   6.8 & 3.6/6.8      & 598.0 &  41.2 \\
             & 559.7 &   9.0 &              & 641.9 &  26.1 \\
             & 596.5 &  19.3 & 5.3/-5.1     & 574.0 &  30.4 \\
             & 624.7 &  16.7 &              & 628.5 &  70.6 \\
             & 647.7 &  10.6 & 5.3/-3.4     & 580.4 &  31.5 \\
-3.2/6.8     & 527.6 &   5.0 &              & 631.0 &  42.2 \\
             & 599.7 &  34.9 & 5.3/-1.7     & 618.5 &  57.9 \\
-1.5/-5.1    & 588.4 &  53.0 & 5.3/0        & 614.1 &  52.7 \\
             & 636.0 &  30.3 & 5.3/1.7      & 568.1 &   5.0 \\
-1.5/-3.4    & 550.5 &  15.8 &              & 608.1 &  49.0 \\
             & 615.2 &  54.7 & 5.3/3.4      & 572.9 &   5.0 \\
             & 675.7 &  10.8 &              & 601.0 &  59.9 \\
-1.5/-1.7    & 529.4 &   5.5 & 5.3/5.1      & 589.7 &  47.2 \\
             & 612.6 &  29.5 &              & 677.7 &   5.0 \\
             & 644.4 &  18.4 & 5.3/6.8      & 587.7 &  61.8 \\
\hline 
\hline
\end{tabular}
\end{table}

\begin{table}[htb]
\caption{Gaussian fit results for NGC3727S}
\label{fits_s}
\begin{tabular}{lrr}
\hline 
\hline
Offset & $v_{\rm{peak}}$ & $\Delta v$ \\
$('')$ & (km\,s$^{-1}$) & (km\,s$^{-1}$) \\
\hline
-5.6/-4.2 &  906.8 &   19.3 \\
-5.6/-2.5 &  903.8 &   26.8 \\
-5.6/-0.8 &  896.0 &   37.6 \\
-5.6/0.9  &  849.5 &   64.0 \\
          &  907.8 &   31.3 \\
-5.6/2.6  &  861.5 &   47.7 \\
          &  906.5 &   31.3 \\
-5.6/4.3  &  874.8 &  107.8 \\
-3.9/-4.2 &  904.5 &   23.5 \\
-3.9/-2.5 &  899.7 &   25.9 \\
-3.9/-0.8 &  820.7 &   38.9 \\
          &  895.8 &   35.2 \\
-3.9/0.9  &  843.5 &   74.8 \\
          &  897.6 &   39.2 \\
-3.9/2.6  &  884.8 &   69.8 \\
-3.9/4.3  &  825.3 &    7.7 \\
          &  884.5 &   65.9 \\
-2.2/-4.2 &  903.8 &   28.7 \\
-2.2/-2.5 &  900.2 &   33.3 \\
-2.2/-0.8 &  819.0 &   22.8 \\
          &  892.4 &   48.4 \\
-2.2/0.9  &  826.5 &   67.0 \\
          &  902.1 &   36.7 \\
-2.2/2.6  &  886.4 &   74.3 \\
-2.2/4.3  &  884.5 &   64.2 \\
-0.5/-4.2 &  821.6 &   14.2 \\
          &  852.6 &   22.8 \\
          &  908.2 &   37.1 \\
-0.5/-2.5 &  902.5 &   30.0 \\
-0.5/-0.8 &  832.0 &   53.1 \\
          &  891.1 &   34.3 \\
-0.5/0.9  &  813.3 &   29.7 \\
          &  887.5 &   56.8 \\
-0.5/2.6  &  893.2 &   58.8 \\
-0.5/4.3  &  817.5 &   54.7 \\
          &  886.8 &   59.0 \\
1.2/-4.2  & --     &        \\
1.2/-2.5  &  884.4 &   71.6 \\
1.2/-0.8  &  838.0 &   40.4 \\
          &  891.4 &   37.4 \\
1.2/0.9   &  867.3 &   60.7 \\
          &  903.1 &   14.7 \\
1.2/2.6   &  787.5 &   21.1 \\
          &  884.8 &   67.9 \\
1.2/4.3   &  784.3 &   12.6 \\
          &  882.3 &   60.0 \\
2.9/-4.2  &  891.2 &   67.2 \\
2.9/-2.5  &  866.2 &   19.3 \\
          &  906.0 &   27.0 \\
2.9/-0.8  &  844.9 &   59.8 \\
          &  905.7 &   22.0 \\
2.9/0.9   &  839.6 &   33.8 \\
          &  900.4 &   29.3 \\
2.9/2.6   &  847.5 &   41.4 \\
          &  901.0 &   31.9 \\
2.9/4.3   &  845.5 &   36.1 \\
          &  900.7 &   35.3 \\
4.6/-4.2  &  895.2 &   41.6 \\
4.6/-2.5  &  774.8 &   11.4 \\
          &  869.9 &  100.2 \\
4.6/-0.8  &  904.2 &   29.8 \\
4.6/0.9   &  839.3 &   62.8 \\
          &  900.7 &   28.4 \\
4.6/2.6   &  809.9 &    8.1 \\
          &  848.5 &   62.3 \\
          &  893.6 &   18.4 \\
4.6/4.3   &  812.4 &    9.5 \\
          &  848.6 &   52.7 \\
          &  916.6 &   13.7 \\
\hline 
\hline
\end{tabular}
\end{table}

\end{document}